\documentclass{desyproc}

\graphicspath{{figures/}}

\usepackage{slashed,mathrsfs,bm,graphicx}

\newcommand{\lagr}{\mathscr{L}}

\newcommand{\ovr}{\overline}
\newcommand{\pp}{\Phi}
\newcommand{\PP}{\Phi^\dagger}

\begin{document}

\title{A global study of the extended scalar singlet model}

\author{{\slshape Ankit Beniwal$^{1,2}$\footnote{Speaker}~, Marek Lewicki$^{2,3}$, Martin White$^2$ and Anthony G. Williams$^2$} \\[1ex]
$^1$The Oskar Klein Centre for Cosmoparticle Physics, Department of Physics, Stockholm University, AlbaNova, SE-106 91 Stockholm, Sweden \\[0.5ex]
$^2$ARC Centre of Excellence for Particle Physics at the Terascale (CoEPP) and CSSM, Department of Physics, University of Adelaide, South Australia 5005, Adelaide, Australia \\[0.5ex]
$^3$Kings College London, Strand, London, WC2R 2LS, United Kingdom}

\contribID{Beniwal\_Ankit}

\confID{20012}  
\desyproc{DESY-PROC-2018-03}
\acronym{Patras 2018} 
\doi  

\maketitle

\begin{abstract}
We present preliminary results from a global study of the extended scalar singlet model with a fermionic dark matter (DM) candidate. In addition to requiring a successful electroweak baryogenesis, we combine constraints from the DM relic density, direct detection limits from PandaX-II experiment, electroweak precision observables and Higgs searches at colliders. In agreement with previous studies, we find that the model can simultaneously explain (at least a part of) the observed DM abundance and matter-antimatter asymmetry. The viable points often lead to strong gravitational wave (GW) signals that can potentially be probed at future GW experiments.
\end{abstract}

\section{Singlet fermion dark matter model}
We extend the Standard Model (SM) by adding a new real scalar singlet $S$ and a Dirac fermion dark matter (DM) field $\psi$. The model Lagrangian is given by \cite{Beniwal:2018pg}
\begin{equation}\label{eqn:model_lagr}
    \lagr = \lagr_{\textnormal{SM}} + \lagr_S + \lagr_\psi + \lagr_{\textnormal{portal}},
\end{equation}    
where $\lagr_{\textnormal{SM}}$ is the SM Lagrangian,
\begin{align}
    \lagr_S &= \frac{1}{2} (\partial_\mu S) (\partial^\mu S) + \frac{1}{2} \mu_S^2 S^2 + \frac{1}{3} \mu_3 S^3 - \frac{1}{4} \lambda_S S^4, \label{eqn:S-lagr} \\
    \lagr_\psi &= \ovr{\psi} (i \slashed{\partial} - \mu_{\psi})\psi - g_S \ovr{\psi} \psi S, \label{eqn:psi-lagr} \\
    \lagr_{\textnormal{portal}} &= -\mu_{\pp S} \PP \pp S - \frac{1}{2} \lambda_{\pp S} \PP \pp S^2. \label{eqn:portal_lagr}
\end{align}
A linear term of the form $\mu_1^3 S$ is removed by a constant shift $S \rightarrow S + \sigma$. When $\mu_3 = g_S = \mu_{\pp S} = 0$, the model reduces to the scalar Higgs portal \cite{Silveira:1985rk,McDonald:1993ex}.

After electroweak symmetry breaking (EWSB), both $\pp$ and $S$ acquire the following VEVs in the unitary gauge  
\begin{equation}
	\pp = 
	\frac{1}{\sqrt{2}}
	\begin{pmatrix}
		0 \\
		v_0 + \varphi
	\end{pmatrix}, 
	\quad 
	S = s_0 + s.
\end{equation}
Consequently, the fermion DM picks up a mass term, i.e., $m_\psi = \mu_\psi + g_S s_0$.

The portal interaction Lagrangian in Eq.~\eqref{eqn:portal_lagr} induces a mixing between $\varphi$ and $s$ fields. To diagonalise the squared mass matrix $\mathcal{M}^2$, we introduce the physical mass eigenstates $(h, H)$ as 
\begin{equation}
    \begin{pmatrix}
        h \\
        H
    \end{pmatrix} = 
    \begin{pmatrix}
        \cos\alpha & -\sin\alpha \\
        \sin\alpha & \cos\alpha
    \end{pmatrix}
    \begin{pmatrix}
        \varphi \\
        s
    \end{pmatrix},
\end{equation}
where $\alpha$ is the mixing angle. Thus, for small mixing, $h$ is a SM-like Higgs boson, whereas $H$ is dominated by the scalar singlet. 

\section{Constraints}


\begin{table}[t]
	\centering
	\begin{tabular}{|c|c|c|c|}
		\hline 
		Constraints (Experiment) & Criteria & Likelihood & Ref. \\\hline 
        Relic density (\emph{Planck}) & $\Omega_\psi h^2 \leq \Omega_{\textnormal{DM}} h^2 = 0.1188$ & one-sided Gaussian & \cite{Ade:2015xua} \\[1mm]
        Direct detection (PandaX-II) & $\sigma_{\textnormal{SI}}^{\textnormal{eff}} \leq \sigma_{\textnormal{PandaX-II}}$ & one-sided Gaussian & \cite{Cui:2017nnn} \\[1mm]
        Electroweak baryogenesis & $v_c/T_c \geq 0.6$ & one-sided Gaussian & --\\[1mm]
        Electroweak precision observables & 
        \begin{tabular}{c}
        $\Delta S = 0.04 \pm 0.11$ \\
        $\Delta T = 0.09 \pm 0.14$ \\
        $\Delta U = -0.02 \pm 0.11$
        \end{tabular}
        & 3D Gaussian & \cite{Haller:2018nnx} \\[1mm]        
        Direct Higgs searches & -- & Step function & \cite{Bechtle:2013wla} \\[1mm]
        Higgs signal strengths & -- & 1D Gaussian & \cite{Bechtle:2013xfa} \\\hline
    \end{tabular}
    \caption{Summary of constraints included in our global fit.~Here $v_c$ is the Higgs VEV at the critical temperature $T_c$.}
    \label{tab:constraints}
\end{table}

In light of the recent discovery of a SM-like Higgs boson \cite{Aad:2012tfa,Chatrchyan:2012xdj}, we set 
\begin{equation}
	m_h = 125.13\,\textnormal{GeV}, \quad v_0 = 246.22\,\textnormal{GeV}.
\end{equation}	
Thus, our model is completely described by the following 7 free parameters
\begin{equation}
	m_H, \quad s_0, \quad \mu_3, \quad \lambda_S, \quad \alpha, \quad m_\psi, \quad g_S.
\end{equation}

To make parameter inferences, we adopt a frequentist approach and perform 7-dimensional (7D) scans of the model using the \texttt{Diver\_v1.4.0} \cite{Workgroup:2017htr} package. The set of constraints included in our global fit are summarised in Table~\ref{tab:constraints}.

\section{Preliminary results}
\begin{figure}[t]
	\centering
	\includegraphics[scale=0.55]{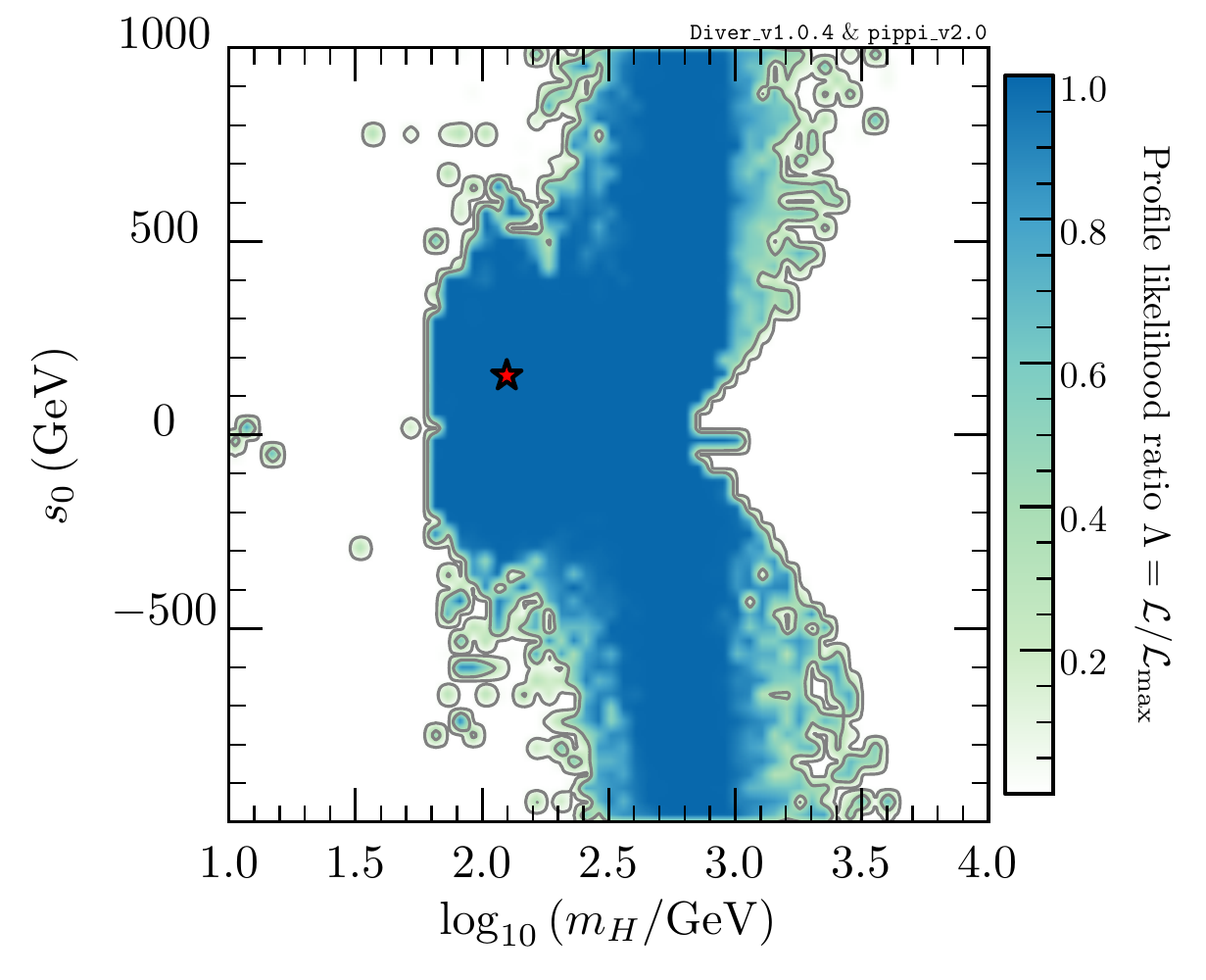} \, 
	\includegraphics[scale=0.55]{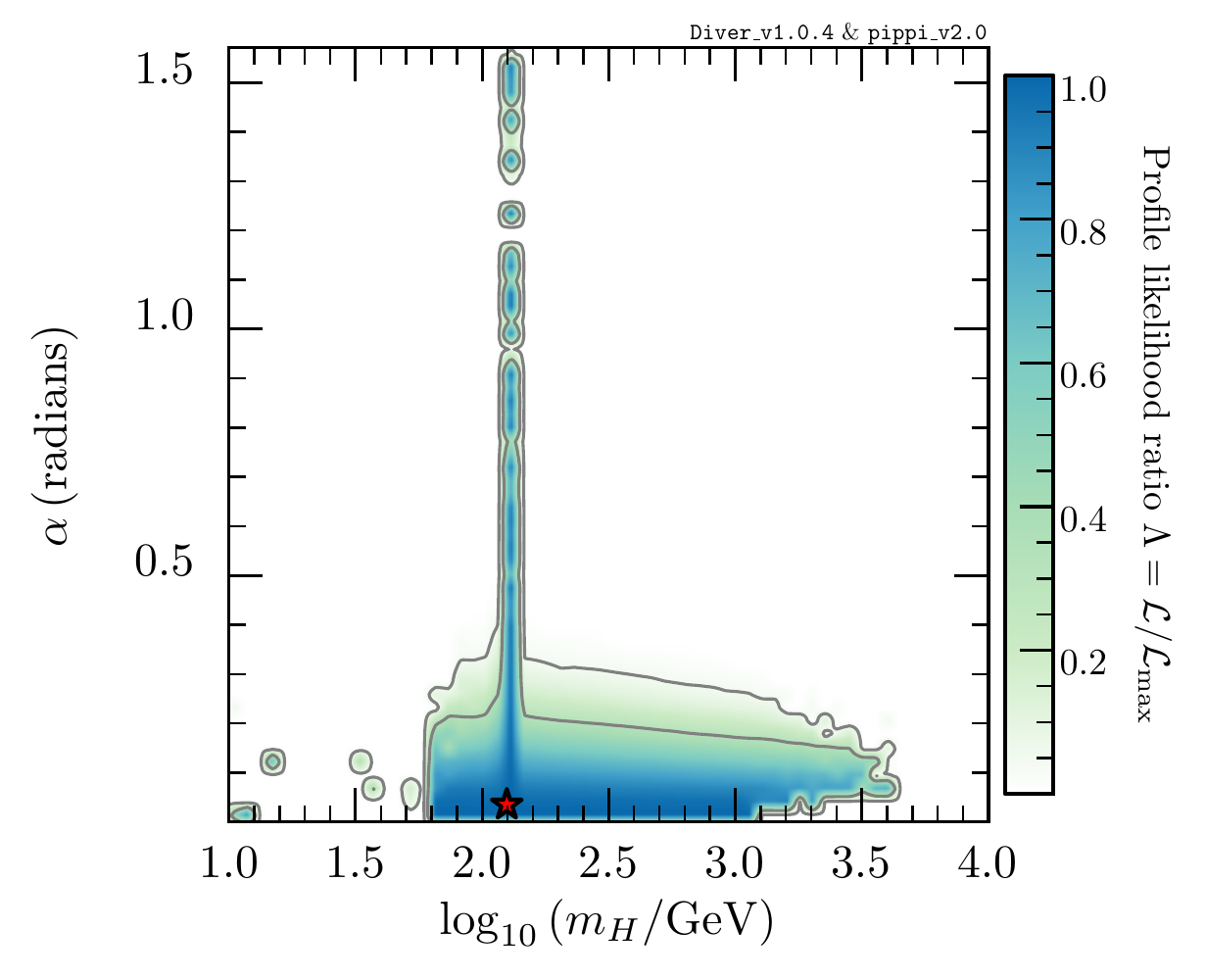}	
	\caption{2D profile likelihood plots from a 7D scan of the model in the $(m_H, s_0)$ and $(m_H, \alpha)$ planes. The best-fit point is marked by the red star.}
	\label{fig:combined_leq}
\end{figure}

In Fig.~\ref{fig:combined_leq}, we show 2D profile likelihood plots from our global fit in the $(m_H, s_0)$ and $(m_H, \alpha)$ planes. Values of $m_H \lesssim m_h/2 = 62.56$\,GeV are ruled out by the observed Higgs signal strengths. This is true regardless of the values of $\alpha$ as the $h \rightarrow HH$ decay mode is dominant in this region, and leads to a reduction of the SM-like Higgs signal strength. When $m_H \simeq m_h$, the model can evade all constraints as the contribution from both scalars cancels out. Thus, all values of $\alpha$ are allowed.~On the other hand, values of $m_H \gtrsim 4$\,TeV are ruled out as they either lead to runaway directions, $\lambda_{\pp S} \leq -2\sqrt{\lambda_\pp \lambda_S}$, or non-perturbative couplings, $|\lambda_{\pp S}|, |\lambda_\pp| \geq 4\pi$.

\begin{figure}[t]
	\centering
	\includegraphics[width=0.8\textwidth]{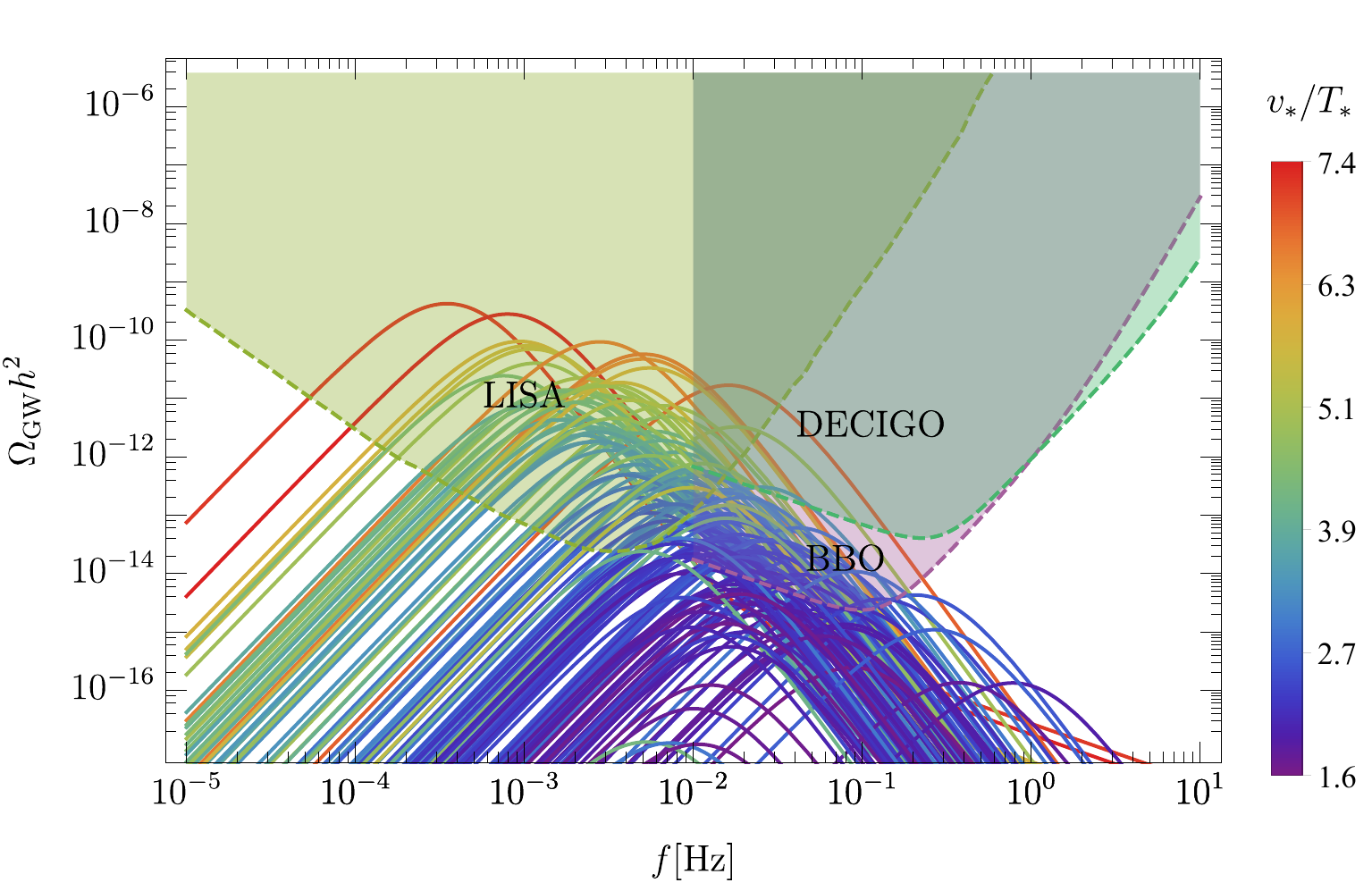}
	\caption{Gravitational wave (GW) spectra of viable points along with the projected sensitivities of future GW experiments such as LISA, DECIGO and BBO. Here $v_*$ is the Higgs VEV at the transition temperature $T_*$.}
	\label{fig:GW_signals}
\end{figure}

For the viable points that satisfy all constraints, we compute their gravitational wave (GW) spectra using the expressions given in Ref.~\cite{Beniwal:2017eik}. The dependence of the GW spectra on $v_*/T_*$ is shown in Fig.~\ref{fig:GW_signals}. For comparison, we also show the detection prospects of future GW experiments such as LISA, DECIGO and BBO. As is evident from the plot, large values of $v_*/T_*$ lead to a stronger phase transition and a stronger GW signal. In particular, they lead to better prospects for detection at future GW experiments.

\section{Conclusions}
We have presented preliminary results from a global fit of the extended scalar singlet model with a fermionic DM candidate. Using the constraints from the \emph{Planck} measured DM relic density, direct detection limits from the PandaX-II experiment, electroweak baryogenesis, electroweak precision observables and Higgs searches at colliders, we performed a global fit of the model. In agreement with previous studies, we found that the model can explain (at least a part of) the observed DM abundance and matter-antimatter asymmetry. In addition, the gravitational wave (GW) spectra of viable points are often found to be within reach of future GW experiments.

\section*{Acknowledgements}
AB thanks the organizers of the Patras 2018 workshop for the opportunity to present this work.~This work was supported by the Swedish Research Council (contract 621-2014-5772), ARC Centre of Excellence for Particle Physics at the Terascale (CoEPP) (CE110001104) and the Centre for the Subatomic Structure of Matter (CSSM). ML was supported in part by the Polish MNiSW grant IP2015 043174 and STFC grant number ST/L000326/1.~AB was supported by the Australian Postgraduate Award (APA). MW is supported by the Australian Research Council Future Fellowship FT140100244.

\begin{footnotesize}

\end{footnotesize}



\begin{thebibliography}{99}
	\raggedright
	\bibitem{Beniwal:2018pg}
	  A.~Beniwal, M.~Lewicki, M.~White and A.~G.~Williams,
	  in preparation for JHEP (2018).	  

	\bibitem{Silveira:1985rk}
	  V.~Silveira and A.~Zee,
	  Phys.\ Lett.\  {\bf 161B} (1985) 136.
	  doi:10.1016/0370-2693(85)90624-0.
	
	\bibitem{McDonald:1993ex} 
	  J.~McDonald,
	  Phys.\ Rev.\ D {\bf 50}, 3637 (1994)
	  doi:10.1103/PhysRevD.50.3637
	  [hep-ph/0702143 [hep-ph]].
	  
	\bibitem{Aad:2012tfa}
	  G.~Aad {\it et al.} [ATLAS Collaboration],
	  Phys.\ Lett.\ B {\bf 716} (2012) 1
	  doi:10.1016/j.physletb.2012.08.020
	  [arXiv:1207.7214 [hep-ex]].
	
	\bibitem{Chatrchyan:2012xdj}
	  S.~Chatrchyan {\it et al.} [CMS Collaboration],
	  Phys.\ Lett.\ B {\bf 716} (2012) 30
	  doi:10.1016/j.physletb.2012.08.021
	  [arXiv:1207.7235 [hep-ex]].
	
	\bibitem{Workgroup:2017htr}
	  G.~D.~Martinez {\it et al.} [GAMBIT Collaboration],
	  Eur.\ Phys.\ J.\ C {\bf 77} (2017) no.11,  761
	  doi:10.1140/epjc/s10052-017-5274-y
	  [arXiv:1705.07959 [hep-ph]].
	
	\bibitem{Ade:2015xua}
	  P.~A.~R.~Ade {\it et al.} [Planck Collaboration],
	  Astron.\ Astrophys.\  {\bf 594} (2016) A13
	  doi:10.1051/0004-6361/201525830
	  [arXiv:1502.01589 [astro-ph.CO]].
	
	\bibitem{Cui:2017nnn}
	  X.~Cui {\it et al.} [PandaX-II Collaboration],
	  Phys.\ Rev.\ Lett.\  {\bf 119} (2017) no.18,  181302
	  doi:10.1103/PhysRevLett.119.181302
	  [arXiv:1708.06917 [astro-ph.CO]].
	
	\bibitem{Haller:2018nnx} 
	  J.~Haller, A.~Hoecker, R.~Kogler, K.~Monig, T.~Peiffer and J.~Stelzer,
	  Eur.\ Phys.\ J.\ C {\bf 78}, no. 8, 675 (2018)
	  doi:10.1140/epjc/s10052-018-6131-3
	  [arXiv:1803.01853 [hep-ph]].
	              
	\bibitem{Bechtle:2013wla} 
	  P.~Bechtle, O.~Brein, S.~Heinemeyer, O.~Stal, T.~Stefaniak, G.~Weiglein and K.~E.~Williams,
	  Eur.\ Phys.\ J.\ C {\bf 74}, no. 3, 2693 (2014)
	  doi:10.1140/epjc/s10052-013-2693-2
	  [arXiv:1311.0055 [hep-ph]].
	  
	\bibitem{Bechtle:2013xfa} 
	  P.~Bechtle, S.~Heinemeyer, O.~Stal, T.~Stefaniak and G.~Weiglein,
	  Eur.\ Phys.\ J.\ C {\bf 74}, no. 2, 2711 (2014)
	  doi:10.1140/epjc/s10052-013-2711-4
	  [arXiv:1305.1933 [hep-ph]].
	                  
	\bibitem{Beniwal:2017eik} 
	  A.~Beniwal, M.~Lewicki, J.~D.~Wells, M.~White and A.~G.~Williams,
	  JHEP {\bf 1708}, 108 (2017)
	  doi:10.1007/JHEP08(2017)108
	  [arXiv:1702.06124 [hep-ph]].
\end{thebibliography}
\end{document}